\documentclass[preprint,12pt]{elsarticle}
\usepackage{amssymb}
\usepackage{float}
\usepackage{subfig}
\usepackage{caption}
\usepackage{graphicx}
\usepackage{color}
\usepackage{tabularx}
\usepackage[counterclockwise, figuresleft]{rotating}
\usepackage[hidelinks]{hyperref}
\journal{Journal of Systems and Software}
\begin{document}
\begin{sloppy}
\begin{frontmatter}
\title{Unusual Events in GitHub Repositories}
\author[add1]{Christoph  Treude}
\ead{christoph.treude@adelaide.edu.au}
\author[add2]{Larissa Leite}
\ead{larissaleite@gmail.com}
\author[add3]{Maur\'{i}cio Aniche}
\ead{m.f.aniche@tudelft.nl}

\address[add1]{School of Computer Science, University of Adelaide, Adelaide, Australia}
\address[add2]{Universitat Polit\`{e}cnica de Catalunya, BarcelonaTech, Barcelona, Spain}
\address[add3]{Software Engineering Research Group, Delft University of Technology, Delft, the Netherlands}

\begin{abstract}
In large and active software projects, it becomes impractical for a developer to stay aware of all project activity. While it might not be necessary to know about each commit or issue, it is arguably important to know about the ones that are unusual. To investigate this hypothesis, we identified unusual events in 200 GitHub projects using a comprehensive list of ways in which an artifact can be unusual and asked 140 developers responsible for or affected by these events to comment on the usefulness of the corresponding information. Based on 2,096 answers, we identify the subset of unusual events that developers consider particularly useful, including large code modifications and unusual amounts of reviewing activity, along with qualitative evidence on the reasons behind these answers. Our findings provide a means for reducing the amount of information that developers need to parse in order to stay up to date with development activity in their projects.
\end{abstract}
\end{frontmatter}
\section{Introduction}

As part of their work, software developers create, modify, and delete many artifacts in any given day. While some of these artifacts follow regular patterns (e.g., an issue is closed by a new commit addressing the issue, or a pull request is merged quickly after a few code review comments), others are unusual: A difficult issue might take a particularly long time to address, a controversial pull request might attract an unusually large number of comments, and a disruptive commit might add or delete a lot of files at once.

For any developer participating in a large and active software project, it quickly becomes impossible to stay aware of all commits, issues, or pull requests that are being created or edited. Arguably, it is also not necessary to be aware of all details happening in a codebase or issue tracking system, and tools such as dashboards~\cite{Treude2010} or event feeds~\cite{Fritz2011} have been designed to abstract away some of the details. In addition to the high-level awareness afforded by such tools, other tools have been proposed to bring developers' attention to activities in a project that have the potential of impacting them directly, such as Brun et al.'s Crystal~\cite{Brun2011} or WeCode~\cite{Guimaraes2012} by Guimar\~{a}es and Silva. However, these tools are very specific and provide little information about the project in general.

In a recent study~\cite{Treude2015} investigating the information that developers would like to be kept aware of, \textit{unusual events} emerged from our qualitative data analysis as a major theme. In fact, we coded 121 out of 156 responses to be related to unusual events or one of its sub-codes. Our work identified a few anecdotal examples of such unusual events, namely an unusually long time between commits by a particular developer, an unusual commit message, or changes to a large number of files. Based on the answers (examples of unusual events that developers are interested in), we hypothesize that developers are interested in unusually large or small values for commit- and issue-related metrics (by generalizing the examples). In this work, we provide a systematic empirical investigation of the hypothesis that developers want to be kept aware of such events in their repositories.

Given the amount of data available in repositories on hosting sites such as GitHub, there is a large number of ways in which an artifact can be unusual. For example, a commit might delete an unusually large number of lines of code, an issue might have an unusually large number of labels, or a pull request might have an unusually large number of commits associated to it. In fact, in this work we found that more than half of all commits in a sample of 200 GitHub projects could be considered as unusual according to at least one metric, considering a comprehensive list of metrics that we defined based on previous work and the data available through the GitHub API.

However, we do not claim that all the different ways in which an artifact could be considered as unusual provide useful information to developers. In contrast, the goal of this work is to enumerate the subset of unusual events that developers consider useful to be kept aware of and to identify the reasons why some types of unusual events are useful to know about and others are not. We define an \textit{unusual event} as an artifact that is unusual in at least one way (e.g., a commit with an unusually large number of files added), and an \textit{unusual event type} as one way in which an artifact could be considered unusual (e.g., unusually large number of files added in a commit). One artifact could be unusual according to more than one unusual event type at any point in time. In this work, we consider commits, issues, and pull requests as artifacts, since they are the main artifacts on GitHub capturing developer activity.

To achieve our research goal of identifying the set of unusual event types that developers consider useful to be kept aware of, we presented 140 developers from 200 randomly sampled GitHub projects with a list of unusual events we had detected in their projects and asked them to rate the usefulness of the corresponding information. Based on a total of 2,096 ratings of different unusual events by the developers that were directly responsible for and/or affected by these unusual events and their reasoning, we compiled a list of types of unusual events that developers want to be kept aware of.

In particular, we investigated the following research questions:

\begin{description}
 \item[RQ1] How are unusual events perceived by developers?
 \begin{description}
 \item[RQ1.1] Are unusual events perceived differently by developers?
 \item[RQ1.2] How do developers perceive artifacts affected by particular types of unusual events?
 \end{description}
 \item[RQ2] Which types of unusual events do developers find most useful and why?
 \begin{description}
 \item[RQ2.1] Which types of unusual events do developers find most useful?
 \item[RQ2.2] Why do developers consider types of unusual events useful or not useful, respectively?
 \end{description}
\end{description}

We found that information on unusual events in terms of number of lines of code deleted, added, and modified in a commit was considered particularly useful, along with the number of comments on issues and pull requests as well as the duration for which an issue had been open. These are also the types of unusual events that belong to artifacts perceived as difficult.

The contributions of this work are:

\begin{itemize}
\item a list of types of unusual events that developers want to be kept aware of, based on empirical evidence, 
\item the reasons for including and excluding specific unusual event types from this list, 
\item data from 200 randomly sampled GitHub projects about the frequency of unusual events and their types, and 
\item an investigation to what extent different types of unusual events correlate with perceived difficulty and typicality of an artifact.
\end{itemize}

The remainder of this paper is structured as follows: Section~\ref{sec:motivating} provides motivating examples for this work. In Section~\ref{sec:projects}, we detail our sampling method for GitHub projects and we provide our definition of unusual events. Section~\ref{sec:frequency} provides empirical data on how frequently the various unusual events occur in GitHub projects. Section~\ref{sec:method} presents our research questions and methodology, before Section~\ref{sec:findings} presents the findings which are discussed in Section~\ref{sec:discussion}. Section~\ref{sec:limitations} highlights the limitations, and Section~\ref{sec:related} summarizes related work. We conclude the paper and outline future work in Section~\ref{sec:conclusion}.

\section{Motivating Examples}
\label{sec:motivating}

RxSwift\footnote{\url{https://github.com/ReactiveX/RxSwift}} is a GitHub project that ports ReactiveX, an API for asynchronous programming with observable streams, to Swift. When we downloaded its data, the repository contained 1,605 commits, 352 issues, and 443 pull requests. A typical issue on RxSwift is closed after being open for less than 5 days (median: 4.65 days, first quartile: 21.74 hours, third quartile: 16.20 days). Considering these numbers, issue \#206 is unusual: more than 10 weeks passed between the moment it was opened and the moment it was closed. When we pointed this out to one of RxSwift's contributors, they stated: \textit{``I think the info is really useful actually, having a long standing issue could [...] be an indicator of a difficult issue''}.

Another project we analyzed for this work is LaTeXML,\footnote{\url{https://github.com/brucemiller/LaTeXML}} a converter for \LaTeX~to XML, HTML, and other formats. The corresponding repository contained 4,520 commits, 675 issues, and 119 pull requests when we downloaded its data. Out of the 675 issues, 21 were labeled with \textit{wontfix}. These issues usually did not attract much discussion: the median number of comments for these 21 issues was 2, with the first quartile at 1 and the third quartile at 3.5. Issue \#724 is unusual in this regard with 13 comments. When we asked one of LaTeXML's contributors about this unusual event, they responded: \textit{``In this case it indicates an interesting discussion that spans beyond the concrete issue''}.

Finally, the Elixir repository\footnote{\url{https://github.com/elixir-lang/elixir}} on GitHub hosts a dynamic, functional language for building scalable and maintainable applications, with 11,548 commits, 2,402 issues, and 2,696 pull requests at the time of our data download. Issue \#3413 is unusual in terms of time between open and closed with a duration of almost 11 months, considering all issues in this project assigned to GitHub user josevalim. This user typically closes issues in less than 7 days (median: 6.94 days, first quartile: 21.26 hours, third quartile: 36.21 days). Given these numbers, one of his colleagues commented: \textit{``This information is useful. Knowing Jos\'{e} [...] closes issues quickly makes it appear that this was a difficult problem''}.

The goal of our work is to provide developers with useful insights such as the ones illustrated in these examples through a systematic investigation of different types of unusual events and their perceived usefulness.

\section{Projects and definition of unusual}
\label{sec:projects}

In this section, we explain our method for sampling GitHub projects and the definition of unusual used in this work.

\subsection{Project Selection}

To systematically investigate awareness of unusual events, we randomly selected 200 original projects (excluding forks) from GitHub, limiting our sample to those projects that had at least 500 commits and at least 100 pull requests or 100 issues. The threshold of 500 commits had been used in previous work (e.g.,~\cite{Aniche2016a}) to filter out pet projects and small experiments developers host on GitHub. The additional filter on the number of issues and pull requests ensures that projects use at least one of these mechanisms to manage their development work.

To conduct the project selection, we randomly selected GitHub projects from the entire population of GitHub projects until we had 200 projects that fulfilled our criteria.\footnote{We performed the random selection by randomly selecting GitHub project IDs between 1 and 70,000,000 and testing whether the corresponding projects fulfilled our sampling criteria.} During this process, we disregarded 129,860 projects because they did not have enough commits and not enough issues or pull requests, 118,873 projects because they were forks, 1,335 projects because they did not have enough issues or pulls (but enough commits), and 350 projects because they did not have enough commits (but enough issues or pull requests). In addition, we disregarded 94 projects that had been imported to GitHub using GoogleCodeExporter (i.e., their issue information was from Google Code rather than GitHub), we disregarded one project because it was a book writing project rather than a software project, and we disregarded one project because it was used for coordinating a political campaign rather than software development. The final list of repositories is available in our online appendix.\footnote{\url{http://tinyurl.com/unusual-events-github}}

\begin{table}[]
\centering
\caption{Descriptive statistics of the 200 GitHub projects}
\label{tab:project-stats}
\begin{tabular}{lrrr}
\hline
 & commits & issues & pull requests \\
\hline
total & 426,133 & 65,008 & 56,037 \\
minimum & 509 & 0 & 0 \\
first quartile & 718 & 67 & 58 \\
median & 1,352 & 171 & 142 \\
third quartile & 2,436 & 352 & 267 \\
maximum & 13,004 & 6,340 & 4,126 \\
\hline
\end{tabular}
\end{table}

Table~\ref{tab:project-stats} shows descriptive statistics of the 200 randomly sampled GitHub projects. They account for a total of more than 425,000 commits, with the most active project having more than 13,000 commits and the median project having 1,352 commits. The projects also account for roughly 60,000 issues and pull requests, with medians of 171 and 142, respectively.

\begin{table}[]
\centering
\caption{Number of unusual commits. The column called ``project'' shows how many unusual events we identified using the entire project as context, whereas the remaining columns show the number of unusual events created by context-specific types.}
\label{tab:commit-outliers}
\begin{tabular}{lrrrr}
\hline
 & project & label & merge? & committer \\
\hline
days between commits & 50,333 & 6,963 & 43,850 & 45,387 \\
\textit{} & \textit{12\%} & \textit{2\%} & \textit{10\%} & \textit{11\%} \\
number of LOC added & 40,528 & 5,924 & 38,758 & 36,055 \\
\textit{} & \textit{10\%} & \textit{1\%} & \textit{9\%} & \textit{8\%} \\
number of LOC deleted & 45,553 & 6,986 & 44,286 & 41,223 \\
\textit{} & \textit{11\%} & \textit{2\%} & \textit{10\%} & \textit{10\%} \\
number of LOC modified & 40,572 & 6,142 & 38,858 & 36,289 \\
\textit{} & \textit{10\%} & \textit{1\%} & \textit{9\%} & \textit{9\%} \\
message length & 11,915 & 2,573 & 14,361 & 9,803 \\
\textit{} & \textit{3\%} & \textit{1\%} & \textit{3\%} & \textit{2\%} \\
number of comments & 4,876 & 1,103 & 4,873 & 4,242 \\
\textit{} & \textit{1\%} & \textit{0\%} & \textit{1\%} & \textit{1\%} \\
number of files added & 54,521 & 8,018 & 50,629 & 47,022 \\
\textit{} & \textit{13\%} & \textit{2\%} & \textit{12\%} & \textit{11\%} \\
number of files changed & 1,920 & 267 & 1,839 & 1,796 \\
\textit{} & \textit{0\%} & \textit{0\%} & \textit{0\%} & \textit{0\%} \\
number of files deleted & 17,917 & 2,699 & 17,830 & 17,216 \\
\textit{} & \textit{4\%} & \textit{1\%} & \textit{4\%} & \textit{4\%} \\
number of files modified & 32,162 & 5,421 & 31,685 & 30,406 \\
\textit{} & \textit{8\%} & \textit{1\%} & \textit{7\%} & \textit{7\%} \\
number of files renamed & 11,030 & 1,933 & 10,945 & 10,651 \\
\textit{} & \textit{3\%} & \textit{0\%} & \textit{3\%} & \textit{2\%} \\
number of pull requests & 25,515 & 5,406 & 27,445 & 22,679 \\
\textit{} & \textit{6\%} & \textit{1\%} & \textit{6\%} & \textit{5\%} \\
\hline
\end{tabular}
\end{table}

\begin{table}[]
\centering
\caption{Number of unusual commits, when data is grouped by files and filetypes.}
\label{tab:commit-outliers2}
\begin{tabular}{l@{}rrrrrr}
\hline
 & committer/ & file & file/ & file/ & file/ & filetype \\
 & merge? & & merge? & committer & label & \\
\hline
days betw. & 41,701 & 109,406 & 72,657 & 55,245 & 10,143 & 61,742 \\
\ \ commits & \textit{10\%} & \textit{26\%} & \textit{17\%} & \textit{13\%} & \textit{2\%} & \textit{14\%} \\
LOC added & 33,488 & 99,686 & 83,411 & 60,826 & 10,001 & 55,904 \\
\textit{} & \textit{8\%} & \textit{23\%} & \textit{20\%} & \textit{14\%} & \textit{2\%} & \textit{13\%} \\
LOC del. & 39,104 & 108,537 & 92,020 & 67,044 & 10,653 & 66,559 \\
\textit{} & \textit{9\%} & \textit{25\%} & \textit{22\%} & \textit{16\%} & \textit{2\%} & \textit{16\%} \\
LOC mod. & 33,841 & 91,414 & 78,227 & 57,292 & 9,491 & 55,869 \\
\textit{} & \textit{8\%} & \textit{21\%} & \textit{18\%} & \textit{13\%} & \textit{2\%} & \textit{13\%} \\
\hline
\end{tabular}
\end{table}

\begin{table}[]
\centering
\caption{Number of unusual issues. The column called ``project'' shows how many unusual events we identified using the entire project as context, whereas the remaining columns show the number of unusual events created by context-specific types.}
\label{tab:issue-outliers}
\begin{tabular}{lrrrr}
\hline
 & project & label & assignee & owner \\
\hline
body length & 2,795 & 1,897 & 2,720 & 1,482 \\
\textit{} & \textit{4\%} & \textit{3\%} & \textit{4\%} & \textit{2\%} \\
days betw.~open and closed & 4,521 & 2,596 & 4,142 & 1,890 \\
\textit{} & \textit{7\%} & \textit{4\%} & \textit{6\%} & \textit{3\%} \\
number of comments & 1,888 & 1,643 & 1,863 & 1,300 \\
\textit{} & \textit{3\%} & \textit{3\%} & \textit{3\%} & \textit{2\%} \\
number of labels & 2,670 & 2,192 & 3,452 & 1,635 \\
\textit{} & \textit{4\%} & \textit{3\%} & \textit{5\%} & \textit{3\%} \\
number of master branch & 4,367 & 3,579 & 4,098 & 1,863 \\
\ \ commits & \textit{7\%} & \textit{6\%} & \textit{6\%} & \textit{3\%} \\
title length & 391 & 305 & 391 & 306 \\
\textit{} & \textit{1\%} & \textit{0\%} & \textit{1\%} & \textit{0\%} \\
\hline
\end{tabular}
\end{table}

\begin{table}[]
\centering
\caption{Number of unusual pull requests. The column called ``project'' shows how many unusual events we identified using the entire project as context, whereas the remaining columns show the number of unusual events created by context-specific types.}
\label{tab:pull-outliers}
\begin{tabular}{lrrrrr}
\hline
 & project & label & assignee & owner & merge \\
 & & & & & status \\
\hline
body length & 2,434 & 971 & 2,364 & 1,765 & 2,418 \\
 & \textit{4\%} & \textit{2\%} & \textit{4\%} & \textit{3\%} & \textit{4\%} \\
days betw.~open & 5,865 & 1,809 & 5,785 & 3,492 & 5,299 \\
\ \ and closed & \textit{10\%} & \textit{3\%} & \textit{10\%} & \textit{6\%} & \textit{9\%} \\
days betw.~open & 4,144 & 754 & 3,982 & 2,722 & 4,144 \\
\ \ and merged & \textit{7\%} & \textit{1\%} & \textit{7\%} & \textit{5\%} & \textit{7\%} \\
number of changed & 3,966 & 1,521 & 3,839 & 2,239 & 3,824 \\
\ \ files & \textit{7\%} & \textit{3\%} & \textit{7\%} & \textit{4\%} & \textit{7\%} \\
number of code & 5,020 & 2,125 & 5,028 & 2,843 & 4,922 \\
\ \ review comments & \textit{9\%} & \textit{4\%} & \textit{9\%} & \textit{5\%} & \textit{9\%} \\
number of & 2,506 & 679 & 2,356 & 1,716 & 2,617 \\
\ \ comments & \textit{4\%} & \textit{1\%} & \textit{4\%} & \textit{3\%} & \textit{5\%} \\
number of labels & 1,392 & 769 & 1,344 & 770 & 1,290 \\
 & \textit{2\%} & \textit{1\%} & \textit{2\%} & \textit{1\%} & \textit{2\%} \\
number of LOC & 4,976 & 1,850 & 4,869 & 2,981 & 4,911 \\
\ \ added & \textit{9\%} & \textit{3\%} & \textit{9\%} & \textit{5\%} & \textit{9\%} \\
number of LOC & 5,972 & 2,381 & 5,794 & 3,576 & 5,810 \\
\ \ deleted & \textit{11\%} & \textit{4\%} & \textit{10\%} & \textit{6\%} & \textit{10\%} \\
number of master & 3,441 & 917 & 3,484 & 2,736 & 4,795 \\
\ \ branch commits & \textit{6\%} & \textit{2\%} & \textit{6\%} & \textit{5\%} & \textit{9\%} \\
title length & 67 & 20 & 78 & 152 & 89 \\
\textit{} & \textit{0\%} & \textit{0\%} & \textit{0\%} & \textit{0\%} & \textit{0\%} \\
\hline
\end{tabular}
\end{table}

\subsection{Definition of unusual}
\label{sec:outlier-types}

To bootstrap our investigation of the importance of unusual events, we defined a comprehensive list of unusual event types based on our previous preliminary work on unusual events in SVN repositories~\cite{Leite2015}, our past work on awareness~\cite{Treude2010}, productivity metrics~\cite{Treude2015, Lima2015}, other related work~\cite{Alali2008, Bissyande2013, Rozenberg2016}, and the data available through the GitHub API~\cite{Kalliamvakou2014}. It is important to note that the goal of this comprehensive list of unusual event types was not to identify types that we as researchers considered particularly useful, but rather to follow a systematic and inclusive approach that would allow us to ask developers about the usefulness of the different types of unusual events while reducing the bias that we were bringing to this project. The next section lists all unusual event types we considered in this work.

In order not to introduce bias due to different formulas being used for different types of unusual events, we used the same definition of what we consider as unusual for all types: the extreme outlier definition, also used by Alali et al.~\cite{Alali2008}, according to which $x$ is an extreme outlier if $x < Q1 - 3 \cdot IQR$ or $x > Q3 + 3 \cdot IQR$, where $IQR$ denotes the inter-quartile range between the first quartile $Q1$ and the third quartile $Q3$ of the underlying distribution. In other words, a value in a distribution is considered as an outlier if it is either three inter-quartile ranges above the third quartile or three inter-quartile ranges below the first quartile of that distribution.

An important dimension of unusual events is context since what is unusual depends on many factors, including team size, work dynamics, software process, development cycle, domain, product size, criticality, and development model~\cite{Leite2015}. To account for that, some of the types of unusual events that we defined use the complete set of artifacts (i.e., commits, issues, or pull requests) in a project to compute the corresponding distributions while others are context-specific. For example, the types of unusual events related to commit message length take all commits in a project into account and consider a commit as unusual if that commit's message length is extremely short or extremely long, according to the definition of extreme outliers given in the previous paragraph. On the other hand, the unusual event types related to commit message length for a particular committer look at the commits of each committer in a project separately, and consider a commit as unusual if that commit's message length is extremely short or extremely long given the set of commits authored by the particular committer. Since what it unusual depends on the particular project and development team, none of our unusual event types take data from more than one project into account, i.e., all computations are project-specific.

\section{Unusual event types and their frequency}
\label{sec:frequency}

In this section, we present the unusual event types considered in this work along with empirical data on how frequently each type occurred in our sample of 200 GitHub projects.

\subsection{Commit-related types of unusual events}

Tables~\ref{tab:commit-outliers} and~\ref{tab:commit-outliers2} show all types of unusual events we defined for commits along with how many unusual events each type yielded in the set of 200 GitHub projects. The column called \textit{project} in Table~\ref{tab:commit-outliers} shows how many unusual events we identified using the entire project as context, whereas the remaining columns show the number of unusual events created by context-specific types. For example, the first row in Table~\ref{tab:commit-outliers} indicates that 50,333 commits were considered unusual according to their message length when using all commits from each project as context, totaling 12\% of all commits in our data. The next column indicates that 6,963 commits were considered to be unusual when using labels as context. In this case, we split up all commits in our data by project first and then by the labels that had been applied to issues or pull requests linked to each commit.

We used various sources to construct this comprehensive list of types of unusual events: the type related to the \textit{days between commits} was inspired by our previous work on unusual events in SVN repositories where participants mentioned this type of unusual event as a potentially useful piece of information~\cite{Leite2015}. The types regarding \textit{lines of code} (LOC) and \textit{number of files} were inspired by our previous work on developer productivity~\cite{Lima2015} which indicated that code contribution metrics provide meaningful information to developers. Counting the \textit{number of pull requests} that include a particular commit was inspired by the traceability data available through the GitHub API~\cite{Kalliamvakou2014}, and taking \textit{commit message length} into account was inspired by Rozenberg et al.'s Repograms~\cite{Rozenberg2016}. In terms of context, we use \textit{labels} as one way of categorizing commits in an attempt to capture the nature of the task triggering the commit. For example, a large number of added files in a commit related to an enhancement issue might be expected, but the same event could be unusual if it occurred in a commit related to a bug fix~\cite{Leite2015}. We also consider whether the commit is a \textit{merge}, based on its number of parent commits. The \textit{committer} is important since some contributors might regularly perform large commits while others contribute small changes. As Table~\ref{tab:commit-outliers2} indicates, we combined some of the contextual information for cases such as \textit{committer / merge?} to be able to distinguish between merge commits and regular commits of different developers. In addition, we consider the \textit{files} that a commit touched along with their \textit{filetype}. These pieces of contextual information were again inspired by our previous work~\cite{Treude2015} which identified ``time between commits to a particular file'' as a potentially useful piece of information. As Table~\ref{tab:commit-outliers2} shows, we only considered file-level context for some of the commit-related types of unusual events since the remaining ones would not be sensible (e.g., combinations such as ``number of files deleted for a particular file'').

\subsection{Issue-related types of unusual events}

Table~\ref{tab:issue-outliers} shows the types of unusual events we defined for issues. The types of \textit{title length}, \textit{body length}, \textit{number of comments}, and \textit{number of labels} were inspired by the corresponding types of unusual events for commits while the number of \textit{days between} when an issue was opened and when it was closed was inspired by previous work by Bissyand\'{e} et al.~\cite{Bissyande2013} who investigated what they called ``time-to-fix''. The \textit{number of master branch commits} associated to an issue is another type that takes advantage of the traceability between different artifact types afforded by GitHub. In addition to the \textit{project} as a whole, we consider separately the issues that contain a specific \textit{label}, have a specific \textit{owner} (i.e., the person who created the issue), or have a specific \textit{assignee}. 

\subsection{Pull request-related types of unusual events}

The types we defined for pull requests shown in Table~\ref{tab:pull-outliers} largely follow the same logic as the ones considered for commits and issues. In addition to the number of days between when a pull request was opened and when it was closed, we consider the time between when it was opened and when it was \textit{merged}. Since pull requests have source code attached to them, we additionally consider the number of \textit{files changed} as well as the number of \textit{lines of code added or deleted}. GitHub allows for two types of comments on pull requests: general \textit{comments} that are identical to issue comments, and \textit{code review comments} that are comments on a portion of the unified diff associated with the pull request. We consider both types of comments in our definition of types of unusual events for pull requests. In addition to the different kinds of context we consider for issues, we added \textit{merge status} for pull requests. Arguably, pull requests that are merged should be treated differently from those not merged.

\subsection{Overlap between types of unusual events}

As these tables show, we found instances of all types of unusual events in the 200 GitHub projects. In general, file-specific types, such as the number of days between commits to a particular file, account for a large number of unusual events (26\% of all commits), whereas types related to title length of issues or pull requests have a much lower yield (at most 1\% of all issues or pull requests). Artifacts can be unusual in more ways than one. While less than half of the issues (29.81\%) and pull requests (46.22\%) in our data are unusual according to our list of types of unusual events, 58.81\% of all commits in our data are unusual in at least one way, with a maximum of 1,316 types per commit. Note that this number includes many context-specific types of unusual events, such as a different types for each file, filetype, or label.

A tool that detects more than half of all artifacts as unusual is arguably not useful. In the next section, we describe the research method we followed to narrow down the initial comprehensive list of types of unusual events to the much smaller subset that developers consider useful.

\section{Research Method}
\label{sec:method}

In this section, we present our research questions and the methods used for data collection.

\subsection{Research Questions}

Our research is guided by the following research questions:

\begin{description}
 \item[RQ1] How are unusual events perceived by developers?
 \begin{description}
 \item[RQ1.1] Are unusual events perceived differently by developers?
 \item[RQ1.2] How do developers perceive artifacts affected by particular types of unusual events?
 \end{description}
\end{description}

Our first research question explores developers' perceptions of the unusual events we detect. In particular, we are interested in determining whether developers find unusual events difficult or atypical, both of which are motivated by our previous work~\cite{Leite2015} which found preliminary evidence that unusual commits in SVN repositories were considered to be more difficult while results on typicality were inconclusive. Note that our previous work did not define unusual events in a systematic way. If unusual events are perceived differently from artifacts that are not detected as unusual by any of the unusual event types, we can argue that being aware of such unusual events might be useful. 

The investigation of why developers want to be aware of unusual events is the goal of our second research question:

\begin{description}
 \item[RQ2] Which types of unusual events do developers find most useful and why?
 \begin{description}
 \item[RQ2.1] Which types of unusual events do developers find most useful?
 \item[RQ2.2] Why do developers consider types of unusual events useful or not useful, respectively?
 \end{description}
\end{description}
 
This research question aims at filtering the list of types of unusual events down to those that developers consider to be most useful. In addition to identifying this subset, we are interested in the reasons why certain unusual event types are considered useful or not useful, respectively.

\subsection{Data Collection}

Developers' opinions and perceptions are at the heart of our research questions. Therefore, we created research instruments that enabled the collection of data from developers on the various types of unusual events. While general perceptions on different ways of detecting unusual events might be helpful, it is arguably more valuable to ask developers about concrete unusual events that different types would yield in the projects that they are currently working on.

We targeted as participants all developers that had contributed at least one unusual commit in the first half of 2016 to one of the 200 GitHub projects we analyzed in this work. We focused on this subset of developers for our study since we assume that developers would not remember specifics about unusual events from more than six months ago. The developers were contacted in July and August of 2016. Out of the total of 2,634,265 unusual event instances in our dataset, 380,080 happened in 2016, distributed over 176 of the 200 projects.\footnote{We downloaded the relevant data from GitHub on July 18th, 2016.} After discarding developers with invalid email addresses and merging email addresses that linked to the same GitHub profile, we identified 1,549 potential participants that we emailed a link to a dynamic survey instrument. After questions capturing demographic information, we asked each participant about at most 12 different artifacts from their project: 

\begin{itemize}
\item 3 artifacts (one commit, one issue, and one pull request) that \textit{they} had authored that \textit{were not} unusual,
\item 3 artifacts (one commit, one issue, and one pull request) that \textit{they} had authored that \textit{were} unusual,
\item 3 artifacts (one commit, one issue, and one pull request) that \textit{somebody else} on their project had authored that \textit{were not} unusual, and
\item 3 artifacts (one commit, one issue, and one pull request) that \textit{somebody else} on their project had authored that \textit{were} unusual.
\end{itemize}

\begin{table}
\centering
\caption{Survey excerpt---note that there were up to 12 instances of Question 5 and 6 in each survey and that Question 6 only appeared after participants had answered Question~5}
\label{tab:survey}
\begin{tabular}{lp{12cm}}
\hline
1 & In how many different software development projects are you currently an active participant? \textit{(text box)} \\
2 & Is developing software part of your job? \textit{(yes/no)} \\
3 & What is your job title? \textit{(text box)} \\
4 & For how long have you been developing software (in years)? \textit{(text box)} \\
\hline
5 & Take a look at pull request \newline \textit{https://github.com/BristolTopGroup/AnalysisSoftware/pull/201}. Would you consider this to be a typical pull request for this project? Was it unusually difficult? \textit{(4 answer options)} \\
\hline
6 & The following statements about this pull request are true. Which of these statements would you consider useful to be aware of? \\
6a & This pull request is an outlier in terms of \textbf{number of changed files} with a value of \textbf{72}. Most pull requests with these characteristics have values between 2.0 and 13.0 with a median of 6.0. \textit{(checkbox)} \\
6b & This pull request is an outlier in terms of \textbf{number of master branch commits} with a value of \textbf{26}. Most pull requests with these characteristics have values between 2.0 and 6.0 with a median of 3.0. \textit{(checkbox)} \\
6c & Why would you consider this information (not) useful? \textit{(text box)} \\ 
\hline
\end{tabular}
\end{table}

These 12 artifacts were chosen randomly and displayed in random order. For some participants, we were not able to fulfil some of these criteria, for example if their project did not use pull requests. In those cases, the survey contained fewer than 12 artifacts. For each artifact, we asked the participants whether they would consider the artifact to be typical and/or difficult. Questions 5 and 6 in Table~\ref{tab:survey} show examples of the questions asked for one artifact from the AnalysisSoftware\footnote{\url{https://github.com/BristolTopGroup/AnalysisSoftware}} project on GitHub. After the participant had indicated how typical or difficult the artifact was, additional questions appeared asking about the usefulness of up to five types of unusual events, as shown in Table~\ref{tab:survey} (Questions 6a and 6b). The information displayed contained the name of the unusual event type (e.g., ``number of changed files''), the value for the artifact in question, and information about the underlying distribution in terms of the first quartile, the median, and the third quartile. Note that Question 6 was only visible after making a choice on Question 5. If there were more than five ways in which an artifact was detected as unusual, e.g., a pull request that is unusual in terms of number of comments, number of review comments, labels, title length, body length, and days between opening and closing, the list of five types was chosen randomly. In addition, we asked why participants considered the unusual event information to be useful or not useful, respectively. All surveys were generated automatically based on the data collected from the 200 GitHub projects.

\begin{table}
\centering
\caption{Demographics about the 140 survey participants}
\label{tab:demographics}
\begin{tabular}{lrrrrrrl}
\hline
 & min & Q1 & median & Q3 & max & \\
\hline
\begin{tabular}{@{}l@{}}years of\\experience\end{tabular} & 1 & 4 & 10 & 15 & 35 & 
\begin{minipage}{0.2\textwidth}
\includegraphics[height=10mm, width=18mm]{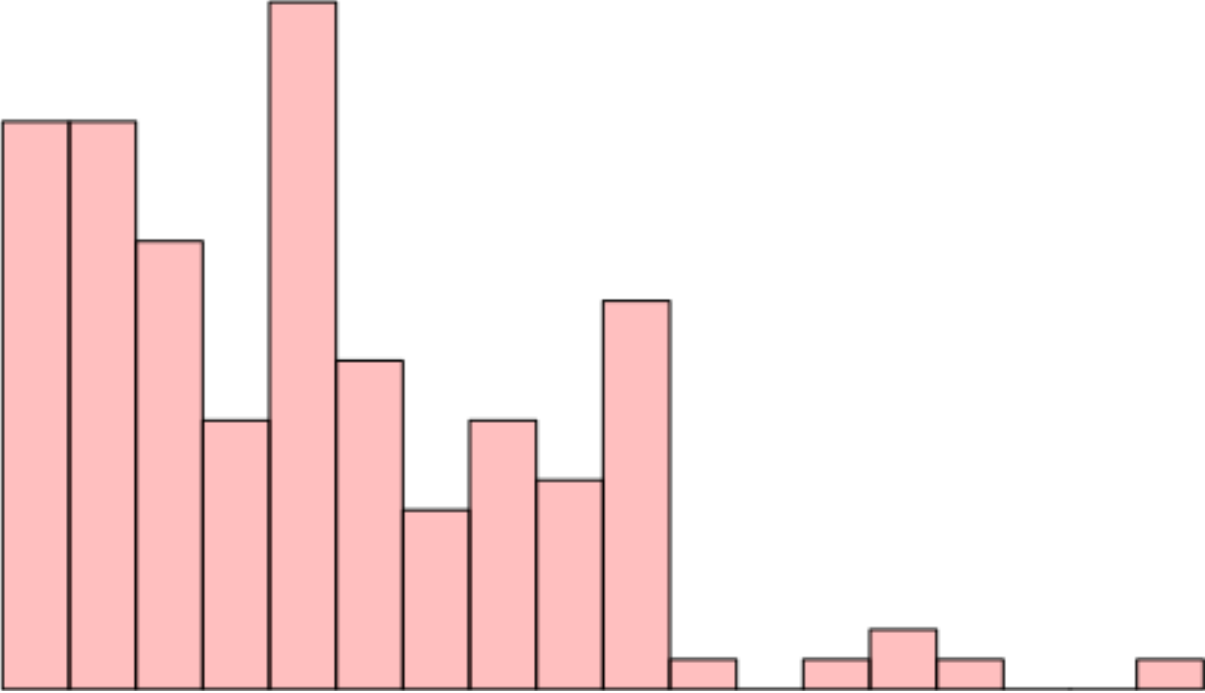}
\end{minipage}
\\[0.5cm]
\begin{tabular}{@{}l@{}}current\\projects\end{tabular} & 0 & 2 & 4 & 5 & 30 & \begin{minipage}{0.2\textwidth}
\includegraphics[height=8mm, width=18mm]{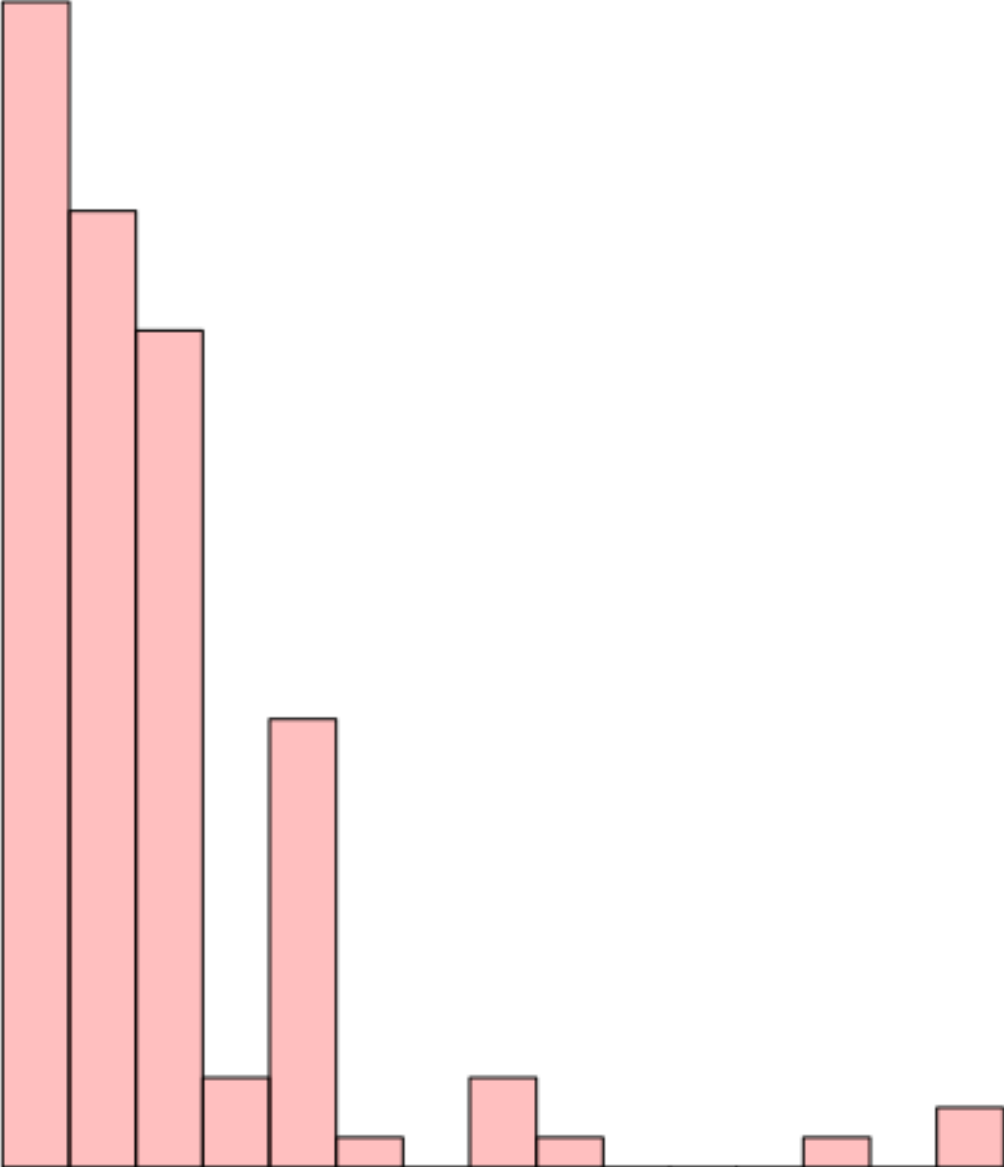}
\end{minipage}
\\[0.3cm] \hline
\end{tabular}
\end{table}

We received 140 responses (response rate: 9.04\%), accounting for a total of 1,157 ratings for different artifacts. In addition to these ratings, we received 2,096 ratings for types of unusual events linked to one of the artifacts and 293 free text answers to the question ``Why would you consider this information (not) useful?''. Most participants were experienced in software development (median: 10 years---3 participants did not answer this question) and involved in more than one project (median: 4---12 participants did not answer this question, see Table~\ref{tab:demographics}). Our participants held different jobs in industry and academia. The majority (66 participants) called themselves ``software developer'' or ``software engineer''. Our sample also contained consultants (3), technical leads (7), directors (3), managers (5), students (11), CEOs or founders (5), and researchers (7).

\section{Findings}
\label{sec:findings}

In this section, we present our findings along with the details on data analysis.

\subsection{Perception of unusual events by developers}

To answer our first research question RQ1.1 on whether unusual events are perceived differently by developers, we analyzed the 1,157 ratings that we received for artifacts. 606 (52\%) of these artifacts were unusual while 551 (48\%) did not have any unusual event type associated to them. These numbers are imbalanced because participants were not required to answer all survey questions and because we were unable to find artifacts with certain characteristics for some of the participants. 436 (38\%) of the artifacts were commits, 305 (26\%) were issues, and 416 (36\%) were pull requests. In addition, 437 (38\%) of the ratings were made on artifacts that the participant had created while 720 (62\%) ratings were made on artifacts that belonged to the same project as the participant, but the participant had not created them. 117 (10\%) artifacts were perceived as being unusually difficult and 317 (27\%) were rated as atypical. 

We calculated the odds ratios and corresponding confidence intervals to analyze whether there was a significant association between whether artifacts were unusual and the ratings they received for difficulty and typicality. In addition to testing across all the ratings in our data, we performed separate tests for artifacts owned by the participant, artifacts not owned by the participant, commits, issues, and pull requests.

\begin{figure}
\centering
\includegraphics[width=\linewidth]{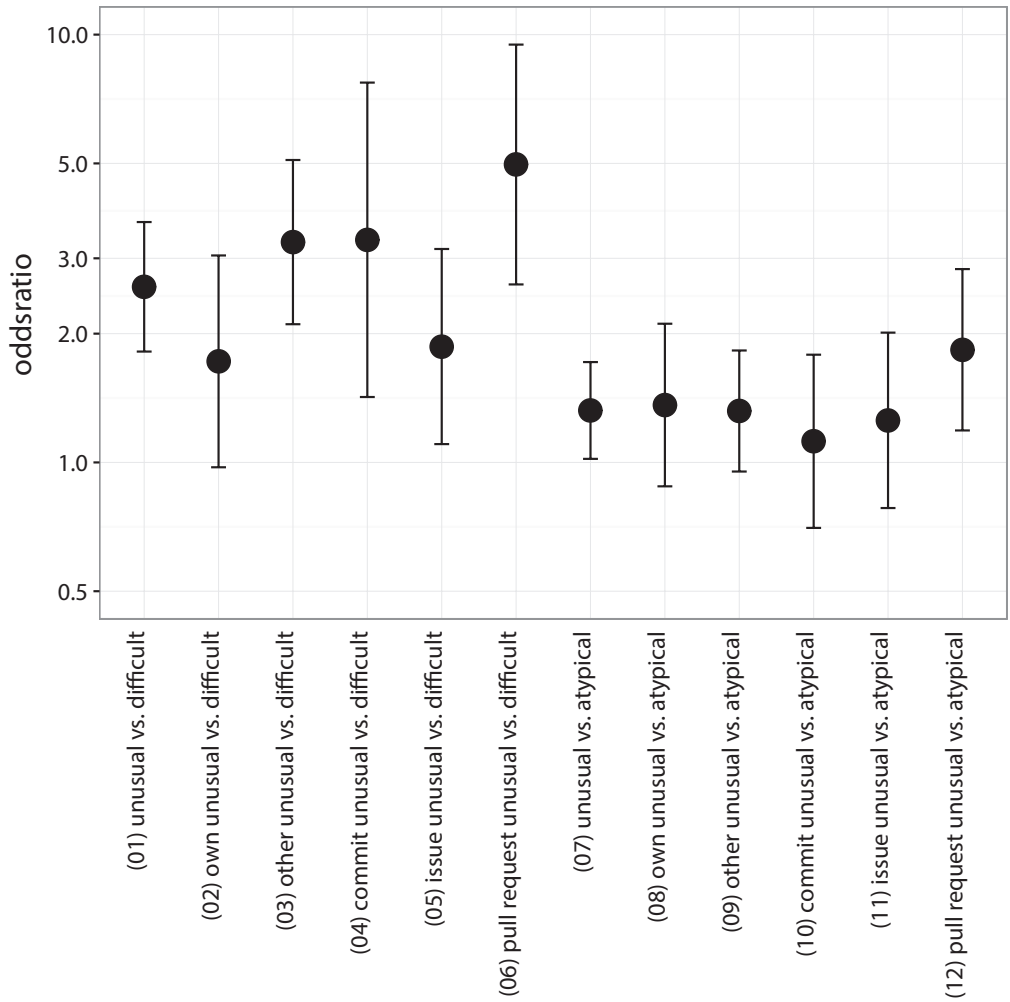}
\caption{Odds of artifacts being perceived as difficult and atypical (log scale). If the 95\% confidence interval does not contain the value 1.0, the association is statistically significant at an alpha level of 0.05.}
\label{fig:oddsratios}
\end{figure}

Figure~\ref{fig:oddsratios} shows the odds ratios along with their 95\% confidence intervals. Confidence intervals wholly to one side of the no effect point (the value 1.0) indicate a significant result~\cite{Altman2013}. The odds of an unusual event being perceived as difficult are 2.57 times higher than the odds of a regular artifact being perceived as difficult (95\% CI [1.81, 3.65]). The odds ratios for artifacts owned by developers other than the participant (3.27, 95\% CI [2.10, 5.09]), commits (3.31, 95\% CI [1.42, 7.72]), and pull requests (4.97, 95\% CI [2.61, 9.48]) are even higher than that. These results are a first indication that one of the use cases of our approach is the detection of difficult artifacts, in particular pull requests.

In contrast, the odds ratios for perceptions of typicality (06--12 in Figure~\ref{fig:oddsratios}) are much lower. None of the odds ratios is higher than 2, and in most cases, the 95\% confidence interval includes values below 1.0, rendering the results inconclusive.

\begin{sidewaysfigure}
 \captionsetup[subfigure]{labelformat=empty}
 \centering
 \subfloat[]{\label{subfig:a} \includegraphics[width=5.8cm]{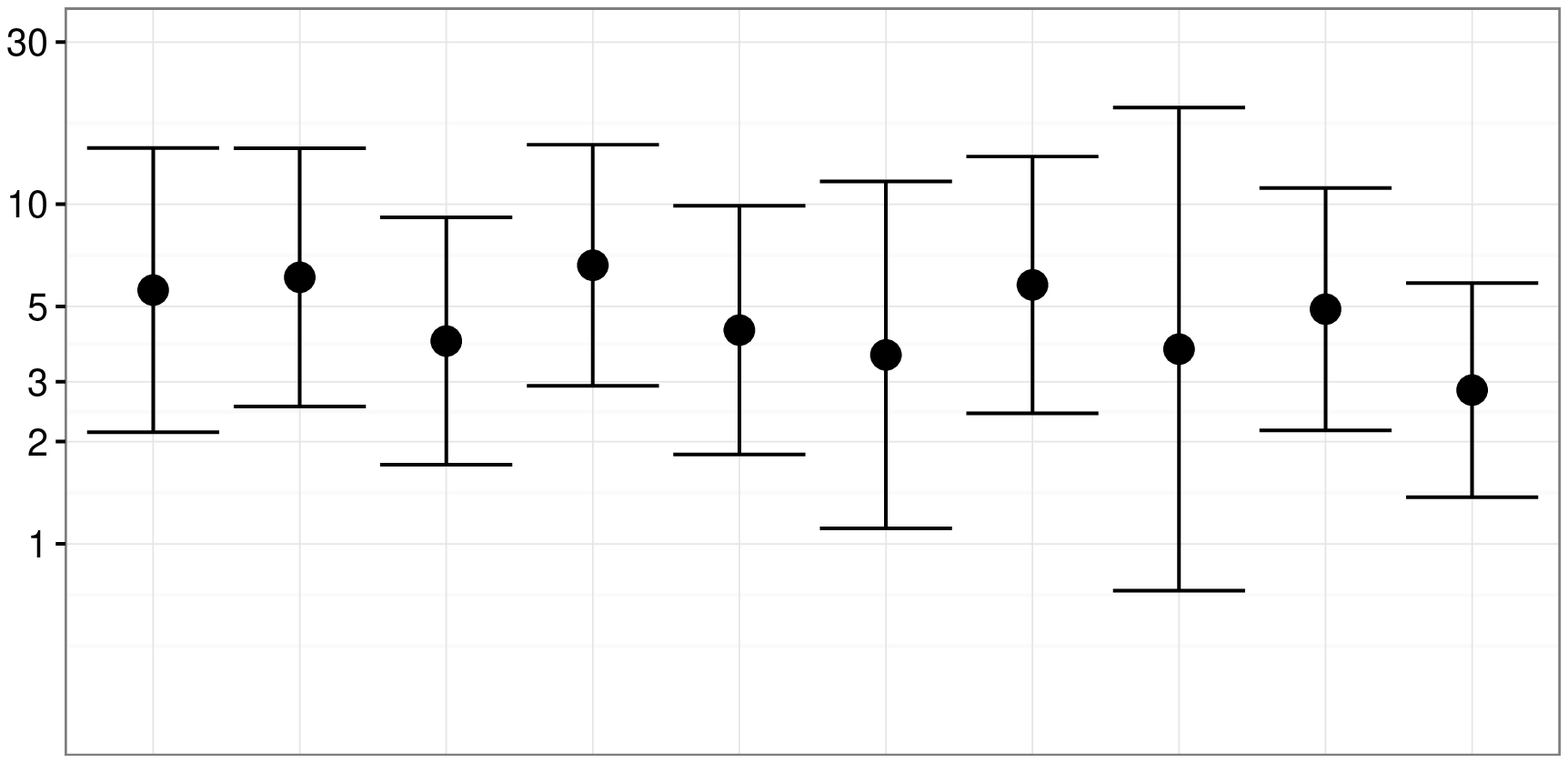}}
 \subfloat[]{\label{subfig:b} \includegraphics[width=5.8cm]{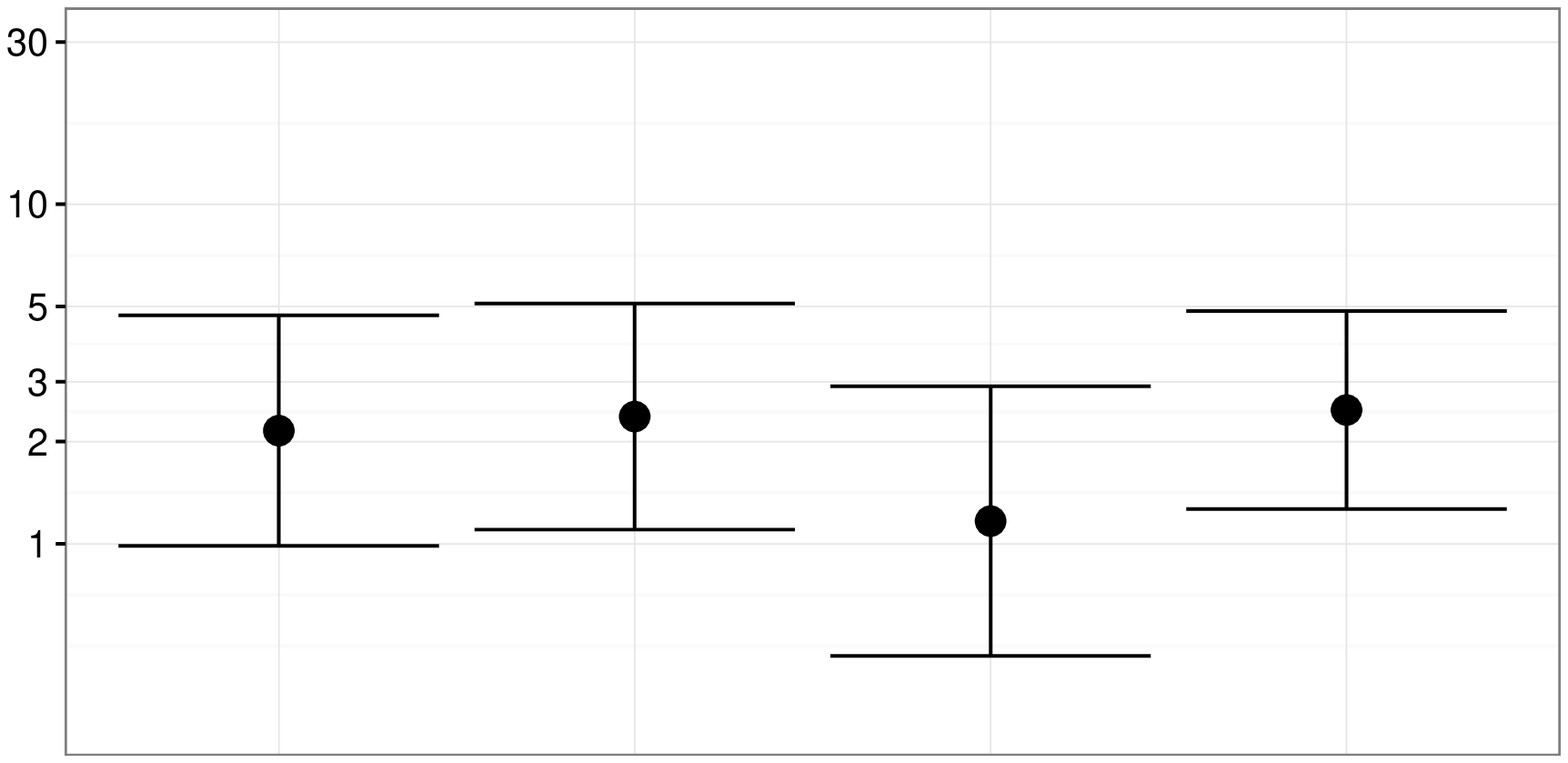}}
 \subfloat[]{\label{subfig:c} \includegraphics[width=5.8cm]{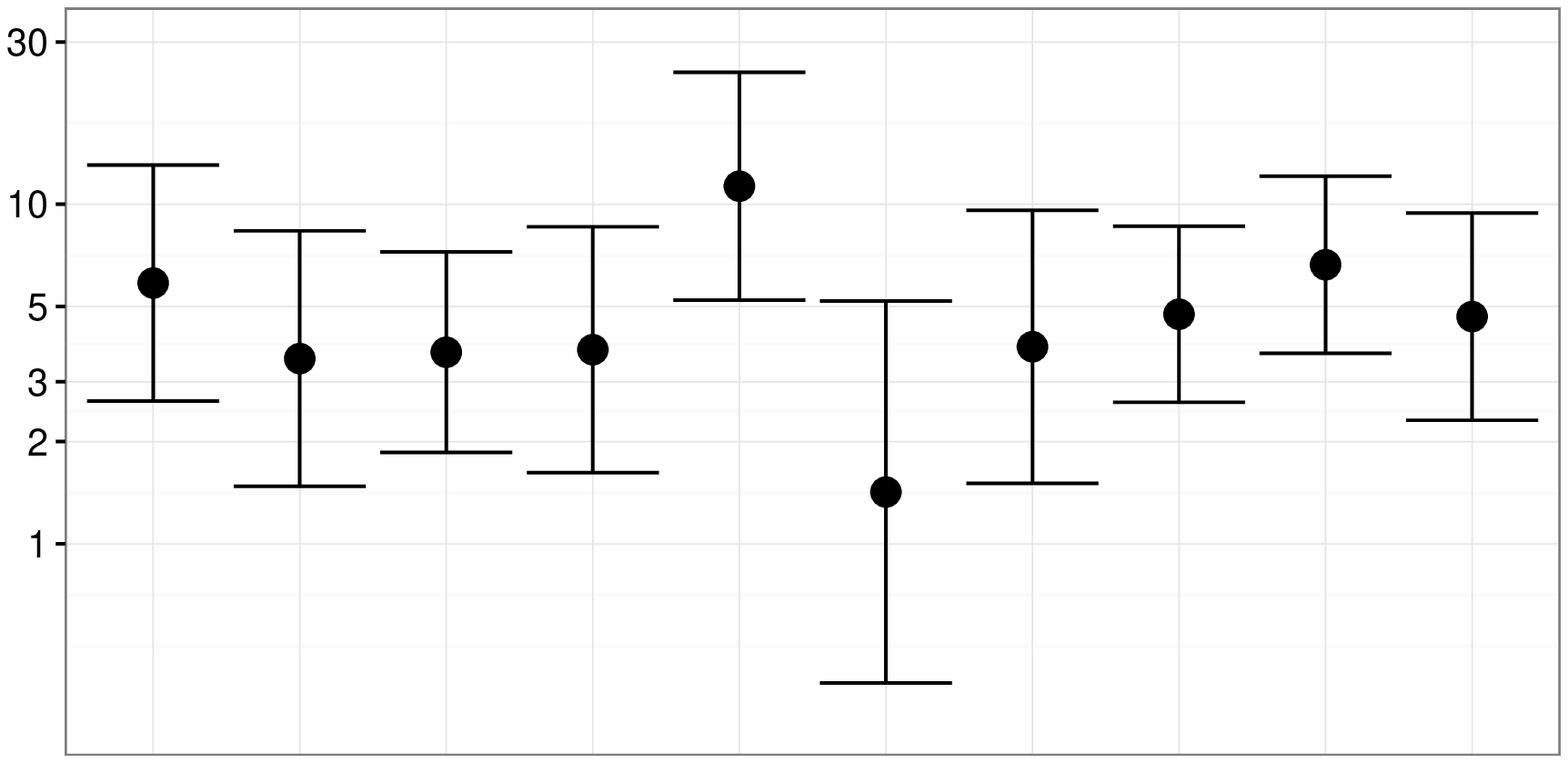}}\\
 \vspace{-2cm}
 \subfloat[]{\label{subfig:d} \includegraphics[trim={0 -1cm 0 0},clip,width=5.8cm]{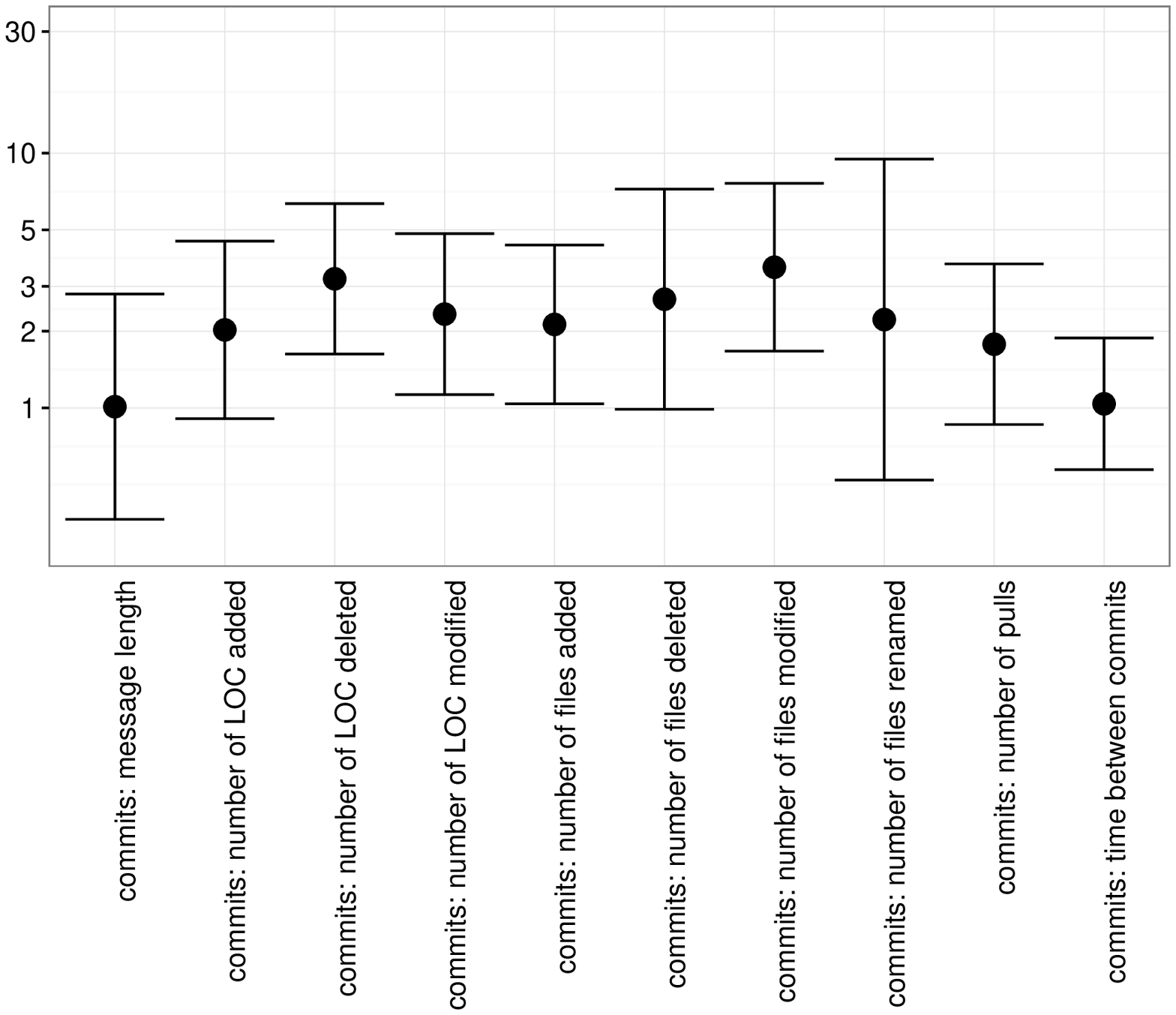}}
 \subfloat[]{\label{subfig:e} \includegraphics[trim={0 -.3cm 0 0},clip,width=5.8cm]{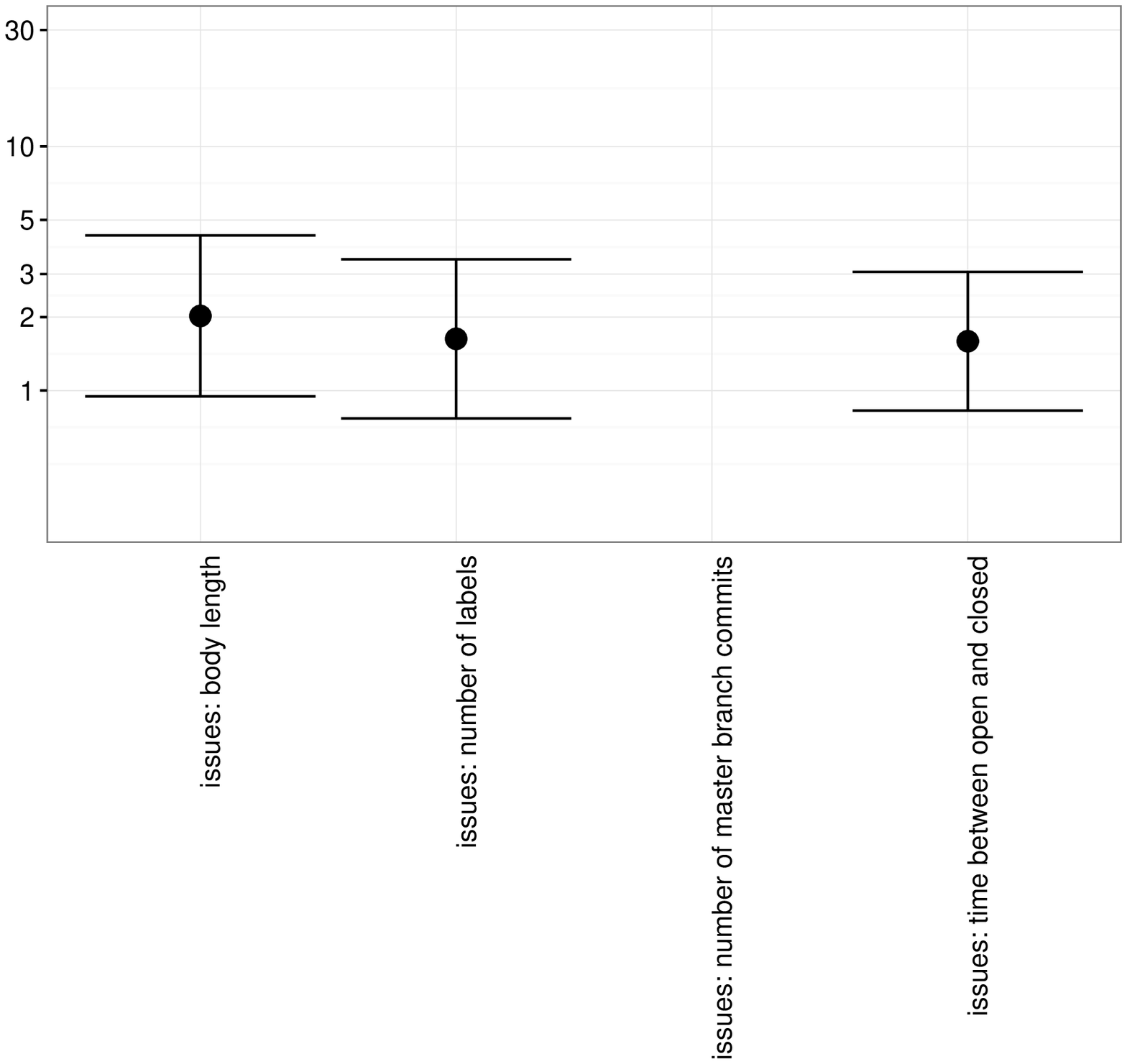}}
 \subfloat[]{\label{subfig:f} \includegraphics[trim={0 -.5cm 0 0},clip,width=5.8cm]{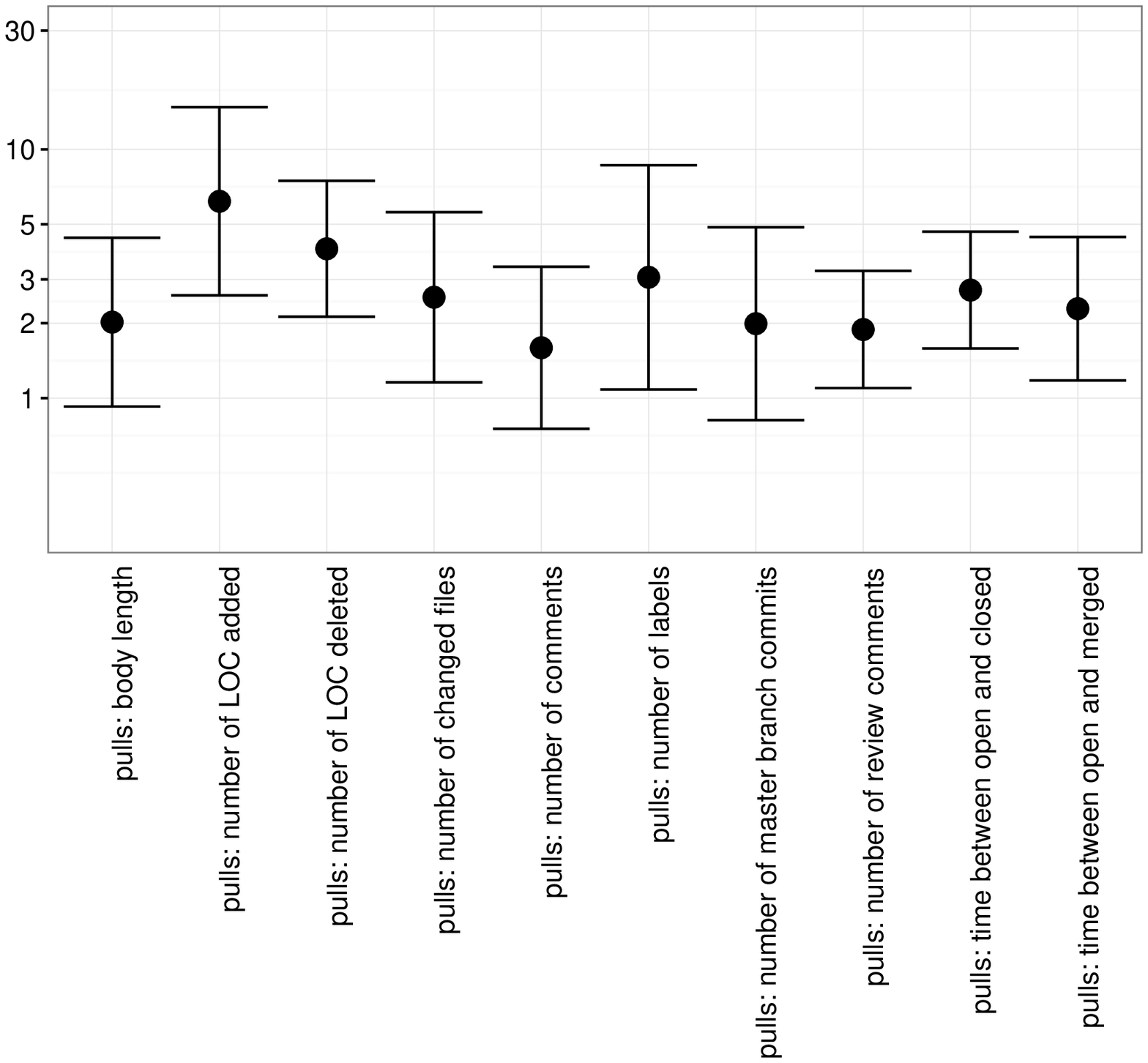}}
 \caption{Odds of artifacts affected by different unusual event types being perceived as difficult and atypical (log scale). The top row shows the odds ratios for an artifact to be considered difficult and the bottom row shows the odds ratios for an artifact to be perceived as atypical. If the 95\% confidence interval does not contain the value 1.0, the association is statistically significant at an alpha level of 0.05. From left to right: commits, issues, and pull requests.}
 \label{fig:difficult-typical-oddsratios}
\end{sidewaysfigure}

To investigate our research question RQ1.2 about developers' perceptions of artifacts affected by particular unusual event types, we calculated the odds ratios for all project-wide unusual event types (second columns in Tables~\ref{tab:commit-outliers} through~\ref{tab:pull-outliers}).\footnote{For context-specific types of unusual events, we did not have enough data to calculate reasonable confidence intervals.} Figure~\ref{fig:difficult-typical-oddsratios} shows the results. The top row shows the odds that an artifact is perceived as being \textit{difficult} if it is affected by an unusual event type compared to the odds of being perceived as difficult if it is not affected by that unusual event type. The bottom row has the corresponding data for artifacts being perceived as \textit{atypical}. The Figure shows unusual event types related to commits, issues, and pull requests from left to right, and we only compare commits to other commits, issues to other issues, and pull requests to other pull requests.

With one exception each, all types of unusual events for commits and pull requests present an odds ratio for difficulty greater than 1.0, even if we take the 95\% confidence interval into account. The highest odds ratio is presented by the number of comments on pull requests: the odds of a pull request that is unusual according to this particular type being perceived as difficult are 11.29 times higher than the odds of other pull requests (95\% CI [5.22, 24.45]). Interestingly, this does not necessarily imply that the pull request is perceived as atypical: As the bottom row of Figure~\ref{fig:difficult-typical-oddsratios} shows, many of the odds ratios for perception of typicality are close to 1.0, with the 95\% confidence interval containing values on either side of 1.0.

Other noteworthy findings include that the number of labels on pull requests does not appear to give information about difficulty (odds ratio: 1.42, 95\% CI [0.39, 5.19]), whereas large pull requests in terms of number of lines of code added are perceived as atypical (odds ratio: 6.18, 95\% CI [2.59, 14.77]). Files being renamed is not a common event, which explains our lack of data and the corresponding large confidence interval for this type of unusual event. A long time between commits is not perceived as atypical (odds ratio: 1.04, 95\% CI [0.57, 1.88]).

\subsection{Usefulness of unusual events}

To investigate our next research question RQ2.1 asking which types of unusual events developers find most useful, we analyzed the 2,096 ratings for the usefulness of different types of unusual events that we received from our participants (see Questions 6a and 6b in Table~\ref{tab:survey}). Our comprehensive list contained 151 types of unusual events (see Section~\ref{sec:outlier-types}), 142 of which received at least one rating in the survey. As a consequence of our comprehensive and inclusive approach to creating the list of types of unusual events, most of the ratings for specific types were negative: we received 414 (19.75\%) positive ratings and 1,682 (80.25\%) negative ratings.

To identify the types of unusual events that received the most positive ratings, we compared the yes/no ratings for each type. Table~\ref{tab:toptypes} shows the six types of unusual events which were rated as useful at least half the time and for which we received more than 5 votes in total. It is unrealistic to assume that developers reach perfect agreement as to what information they find useful. In fact, different stakeholders interested in software repository data have been found to have very different information needs~\cite{Buse2012}.

To understand why some types of unusual events are perceived as more useful than others, two of the authors applied qualitative coding to the 293 answers that participants had given to the question ``Why would you consider this information (not) useful?''. We identified a total of 26 codes. The complete list of answers and assigned codes is available in our online appendix.$^{4}$ In the following, for each code that emerged in the qualitative analysis, we indicate how many participants mentioned the particular theme in superscript and we present a representative quote. Note that the numbers only indicate how much evidence the data analysis yielded for each theme, they do not necessarily indicate the importance of a theme since we did not explicitly ask all participants about each theme specifically.

\begin{table}
\centering
\caption{The most useful types of unusual events}
\label{tab:toptypes}
\begin{tabular}{llr}
\hline
artifact & type & positive votes \\
\hline
commits & number of LOC modified & 60\% \\
& \hspace{.5cm} the number of lines of code modified & \\
& \hspace{.5cm} in a commit & \\
pull requests & number of comments & 57\% \\
& \hspace{.5cm} the number of comments & \\
& \hspace{.5cm} on a pull request & \\
issues & days between open and closed & 53\% \\
& \hspace{.5cm} the time for which an issue was open & \\
issues & number of comments for label & 50\% \\
& \hspace{.5cm} the number of comments on an issue, & \\
& \hspace{.5cm} for a particular issue label & \\
commits & number of LOC deleted & 50\% \\
& \hspace{.5cm} the number of lines of code deleted & \\
& \hspace{.5cm} in a commit & \\
commits & number of LOC added & 50\% \\
& \hspace{.5cm} the number of lines of code added & \\
& \hspace{.5cm} in a commit & \\
\hline
\end{tabular}
\end{table}

As our quantitative data suggests, the number of comments on an issue or pull request is a useful metric$^{(16)}$: \textit{``When reviewing project history, reviews with a lot of comments could be more interesting in terms of decisions''}. While not all participants echoed this opinion, there were only a couple that voiced the opposite$^{(2)}$: \textit{``Comments are not really a good indicator of much in and of them selves''}.

The response time to pull requests and issues is also an important metric$^{(9)}$: \textit{``The responsiveness of the repo owner is valuable when evaluating using/contributing to an open source project''}. A long-standing pull request or issue could also indicate low priority$^{(5)}$: \textit{``It's useful to see we have a long outstanding documentation issue, as they tend to be neglected unless we are reminded accordingly''} or difficulty$^{(1)}$: \textit{``The open-closed time usually means low priority, improper issue statement or something very difficult to fix''}. Seeing which issues have a long time-to-close can be useful$^{(9)}$: \textit{``Yes, seeing which issues have been open the longest can be vital''}, but is less useful after the fact$^{(1)}$: \textit{``It would be good to know which issues have been outstanding for a long time but once the issue has been addressed/closed, this information becomes irrelevant''}.

Another important metric is churn$^{(3)}$: \textit{``It is useful because it can specify the amount the code that is modified, and might even help spot the problem if anything goes wrong in the future''}. In particular knowing about unusual deletions can be useful$^{(3)}$: \textit{``LOC deleted relative to owner can be a good measure of difficulty of the pull request''}, although one participant disagreed$^{(1)}$: \textit{``Core developer deleting one file isn't unusual''}.

Knowing about gaps in the commit activity can also be useful$^{(4)}$: \textit{``Days between commits for filetypes would be good to determine what kinds of things are being worked on (code vs.~assets vs.~build scripts)''}, similar to the number of commits on a pull request$^{(1)}$: \textit{``Detect unusual PRs to have more people check it''}. Types of unusual events related to long commit messages or issue and pull request bodies generally received low ratings$^{(9)}$: \textit{``Commit comment length isn't a useful metric unless its unusually short''}, although some participants saw value$^{(4)}$: \textit{``A longer message can indicate the why for the change is not immediately obvious''}. 

In some cases, the insights extracted by our analysis helped developers reflect on their projects: \textit{``I realized that code review comments rarely happened''} and could possibly be used to gamify some aspects of the development process$^{(5)}$: \textit{``Maybe the typical number of commits a user usually has would be useful for encouraging developers to commit more often''}. This theme is echoed in related work~\cite{Singer2012}.

Missing context was prevalent among the main criticisms of some types of unusual events$^{(5)}$: \textit{``I need to see them in context. Adding 100 lines of documentation probably doesn't need as much attention as adding 100 new functions''}. Commit-related types were often affected by changes to the documentation or formatting$^{(10)}$: \textit{``It is a change in text. Not code. So there are no changes''}, and some commits were generated automatically$^{(4)}$: \textit{``It's a commit made by an automated system. Not interested in getting statistics for these events''}. Gaps in commit activity could be explained by external reasons$^{(8)}$: \textit{``Open source projects are side-time for most of us''}. More advanced code metrics could be useful to address some of these issues$^{(3)}$: \textit{``Displaying a raw complexity score may be useful''}. Most of these limitations can be traced back to our decision to keep the types of unusual events independent of a particular programming language in order to be applicable to all GitHub repositories. Future work will have to investigate this tradeoff further.

In other cases, the information provided by the types of unusual events was too fine-grained$^{(7)}$: \textit{``I think splitting LOC metrics by file and label is probably too granular''}, and unusual events based on labels were generally not seen as useful$^{(4)}$: \textit{``Different projects use GitHub labels for different purposes. Some of them are not using label at all. It's not significant''}.

For types of unusual events outside of the six that we identified as being the most useful, the information was often seen as not useful in a practical sense$^{(26)}$: \textit{``It's interesting but not useful as a contributor''}. In addition, many participants used our survey to explain the unusual events rather than indicate whether the information was useful$^{(78)}$: \textit{``It was just a small improvement to run samples''}, or they only answered whether the information was useful without stating why$^{(66)}$: \textit{``Useful''}, \textit{``Not useful''}.

\section{Discussion}
\label{sec:discussion}

In this section, we discuss our findings, in particular related to the developers' perceptions of unusual events, the events' verifiability, implications for a user interface for displaying unusual events, and actions that developers can take based on unusual events.

\subsection{Perceptions of difficulty and typicality}

When we limit the list of unusual event types to the six types that our participants rated as being the most useful, we identify 15\% of all commits, 8\% of all issues, and 4\% of all pull requests in the 200 GitHub projects used in this study as unusual. Our findings provide evidence that these unusual events are perceived as being particularly difficult by developers and that awareness of such unusual events is considered useful. While developers might be aware of difficult unusual events among their own artifacts, knowing about difficult work of somebody else on the team can help prevent potential problems early on, can encourage discussion where it is needed, and can give important pointers to events in a project's history to be reviewed when trying to understand a project. Interestingly, our participants rated unusual events among their own artifacts as less difficult compared to unusual events among the artifacts of their team members. 

The relationship between the unusual event information and whether an artifact is considered atypical is much weaker, suggesting that there is a difference between observed typicality (via metrics) and perceptions of typicality, and that developers do not view difficult tasks as atypical.

\subsection{Verifiability}

Our findings provide evidence that developers value simple and easily understandable metrics over complex ones. With one exception, the unusual event types that were rated as being most useful are based on project-wide metrics that can easily be verified by looking at the raw data. While our initial set of unusual event types contained more complex and context-specific unusual event types, we found that developers generally did not find these as useful as unusual event types based on project-wide data. In addition, our findings indicate that awareness tools based on commit or source code activity alone are not sufficient to communicate all the information developers care about in a project: half of the six most useful unusual event types are related to issues and pull requests.

We note that while the raw numbers can easily be verified, simply looking at the metrics of a given commit, issue, or pull request will not tell developers whether this artifact is unusual. To acquire this information, developers have to download all data, calculate the metric values for each artifact, and investigate the distributions of the different values---just as we did in this work. In other words, while the raw numbers are verifiable, deciding what is unusual requires a considerable amount of work that developers are unlikely to undertake.

\subsection{User interface}

In terms of user interface, information about unusual events could be integrated into a developer's workflow in various ways: an event feed could be provided on a separate website or notifications could appear in an IDE. A particularly promising approach would be the integration into communication tools through bots~\cite{Lin2016}. For example, the current integration of GitHub events into the cloud-based team collaboration tool Slack creates a notification for each action taken on GitHub. Arguably, it would be more useful to generate these notifications only in cases where something unusual happened that deserves attention. Our empirical investigation of different unusual event types and their perceived usefulness provides the empirical foundation for building such tool support.

\subsection{Implications}

Tool support which surfaces unusual events is useful for software developers and their managers. Unusual events have the potential to significantly reduce the amount of information that developers and managers need to parse to stay on top of everything that is going on in their projects. We found that only 15\% of all commits, 8\% of all issues, and 4\% of all pull requests in our data set were classified as unusual. Having to only look at this small subset of information will save developers' time and make it less likely for them to miss important events that require their attention. Notifications about unusual events can trigger a wide range of actions by software developers or managers. In our previous preliminary work on unusual events in SVN repositories~\cite{Leite2015}, we found that notifications about such events could serve as a discussion starter or a meeting agenda. As one developer explained: \textit{``It would be useful to be aware of unusual events from other developers. [...] If I notice a strange modification or many modifications I can promptly talk to the developer about it.''} Unusual events can also play a significant role for managers who are in charge of monitoring project progress: \textit{``As a manager, it is a way to look closer to what newcomers are doing. [...] The information would be useful in the meetings, since I could question and talk to them about their tasks without being too passive and waiting for them to tell me something''}~\cite{Leite2015}.

Based on these preliminary findings, in this work, we have conducted a systematic exploration of different kinds of unusual events that can be detected for GitHub projects, and we have identified the subset of unusual events that developers find particularly useful. We have uncovered additional use cases for unusual events, for example detailed in Section~\ref{sec:motivating}: An issue with an unusually long time between when it was opened and when it was closed can point to difficulties that might require support from other developers, and an issue with an unusually large number of comments can indicate a discussion that other developers should be aware of. While many artifacts in a GitHub repository can be considered as unusual according to some metric, the goal of our work was to identify those types of unusual events that developers find useful. These findings are also important for researchers who are interested in developer awareness in general or GitHub repositories in particular since they uncover a new category of events that developers care about: unusual events related to commits, issues, and pull requests.

\section{Limitations}
\label{sec:limitations}

In terms of construct validity (i.e., the degree to which a test measures what it claims, or purports, to be measuring), while our initial list of types of unusual events was designed to be comprehensive and based on related work, there could be other important unusual event types that we did not ask our participants about. Our definition of unusual, although based on related work~\cite{Alali2008}, is only one possible way of detecting unusual values in a distribution. Other approaches may have led to different results, and we will continue our empirical investigation into the impact of different definitions on the results. However, our results provide a first systematic and empirical exploration of the idea of unusual events, and without empirical evidence, we cannot determine to what extent other approaches would have resulted in different outcomes. 

In terms of internal validity (i.e., the extent to which a causal conclusion based on a study is warranted), even the most useful unusual event types that we defined still received negative ratings. As our qualitative data shows, it is unrealistic to assume that all developers agree on wanting awareness of the same information. Our qualitative analysis may have introduced bias and error into our interpretation of the developer responses. We mitigated this threat by having two of the authors do the coding.

In terms of external validity (i.e., the extent to which the results of a study can be generalized to other situations and to other people), we cannot generalize our findings to development platforms other than GitHub. However, GitHub now hosts more than 19.4 million active repositories,\footnote{\url{https://octoverse.github.com/}} making it a good starting point for this research. To distribute our survey, we contacted all developers that had contributed at least one unusual commit to one of the 200 projects in our sample within the last six months. However, the 140 individuals who contributed to this study were self-selected volunteers within this sample. The general population on GitHub might have different characteristics and opinions. Thus, we cannot claim that our results generalize to all GitHub users or to the entire population of developers.

\section{Related Work}
\label{sec:related}

Existing work on detecting unusual events in software repositories has mostly focused on detecting specific unusual events, often focusing on bug detection and prevention. Crystal~\cite{Brun2013}, for example, can detect if a developer has not committed for a long time, and if a developer has made changes that conflict with other developers' changes, break the build, or make a test fail. WeCode~\cite{Guimaraes2012} identifies the outcomes of merging all the developers' code at once. We take a broader approach by enumerating a large number of unusual events that can happen in GitHub repositories, and by collecting empirical data about their usefulness. 

There is also a substantial body of work on the detection of buggy commits. Kim et al.~\cite{Kim2008} employed machine learning to determine whether a new software change is more similar to prior buggy changes or prior clean changes. Eyolfson et al.~\cite{Eyolfson2011} found commits submitted between midnight and 4am to be significantly more bug-prone than those submitted at other times, and daily-committing developers to produce less buggy commits. The focus of our work is not on bug detection, but rather on making developers aware of unusual events in their repositories.

The detection of unusual events can be supported by visualizations of the software process~\cite{Hindle2010}, the change history~\cite{VanRysselberghe2004}, or an individual commit~\cite{DAmbros2010}. While some of these allow for the identification of unusual events, they are not as comprehensive as our unusual event types and do not include unusual events on issues or pull requests.

Awareness tools for software developers have historically focused on awareness at source code level. For example, Seesoft~\cite{Eick1992} maps each line of source code to a thin row and uses colours to indicate changes. Augur~\cite{Froehlich2004} extends the idea behind Seesoft by adding software development activities to a Seesoft-style visualization, allowing developers to explore relationships between artifacts and activities. Palant\'{i}r~\cite{Sarma2003, Sarma2006} provides insight into workspaces of other developers, focusing on artifact changes. NeedFeed~\cite{Padhye2014} models code relevance and highlights changes that a developer may need to review. Relevant changes are determined using models that incorportate data mined from a project's software repository. With FASTDash~\cite{Biehl2007}, a developer can determine which team members have source files checked out, which files are being viewed, and what methods and classes are currently being changed. Going beyond source code, WIPDash~\cite{Jakobsen2009} was designed to increase awareness of work items and code activity. Similarly, the dashboard component of IBM's Jazz~\cite{Treude2010} is intended to provide information at a glance and to allow for easy navigation to more complete information. Our work was inspired by the dashboard component in Jazz, but using unusual events as content rather than high-level summaries of artifact counts over time. With its inherent transparency~\cite{Dabbish2012}, GitHub affords group awareness in distributed software development~\cite{Calefato2013, Lanubile2013}, and external websites have started to aggregate data from GitHub~\cite{Singer2013}. Our work adds to this body of knowledge by exploring the concept of unusual event awareness on GitHub.

\section{Conclusions and Future Work}
\label{sec:conclusion}

To investigate our hypothesis that awareness of unusual commits, issues, and pull requests on GitHub is important to developers, we created a comprehensive list of ways in which an artifact on GitHub could be considered unusual. We surveyed 140 developers to capture their perceptions of the unusual events we detected in their projects and the corresponding unusual event information. Based on 2,096 answers, we identified the types of unusual events that developers consider particularly useful, including large code modifications and unusual amounts of comments, along with qualitative evidence on the reasons behind these answers. We found that artifacts affected by unusual events are often perceived as being particularly difficult.

Building on these results, our future work consists of building the tool support that our participants envisioned: a feed of unusual events that developers would like to be kept aware of. Such tool support will enable us to investigate the impact of reducing the amount of information that developers need to parse to stay on top of everything that is going on in their projects, e.g., in the form of field studies. In addition, we plan to investigate the extent to which project and developer characteristics influence the unusual event information that is considered useful by developers, with the goal of building a personalized unusual event awareness feed.

\section*{Acknowledgments}

We thank all developers who participated in our survey for their participation.

\section*{References}

\bibliographystyle{elsarticle-num}
\bibliography{unusual-events-jss}
\end{sloppy}
\end{document}